\documentclass[11pt,a4paper]{article}
\usepackage{jcappub}

\def\bc{\begin{center}}
\def\ec{\end{center}}
\def\b{\begin{equation}}
\def\e{\end{equation}}
\def\ber{\begin{eqnarray}}
\def\eer{\end{eqnarray}}

\def\eg{{\it e.g.~}}
\def \ie {{\em i.e.~~}}

\def \lleq {\lower0.9ex\hbox{ $\buildrel < \over \sim$} ~}
\def \ggeq {\lower0.9ex\hbox{ $\buildrel > \over \sim$} ~}

\def \L {$\Lambda$}

\def \hm1 {$h^{-1}$}
\def \sv {$\sigma^2_v~$}

\title{The multi-stream flows and the dynamics of the cosmic web}
\author{Sergei F.Shandarin}
\affiliation{Department of Physics and Astronomy, University of Kansas, \\
10082 Malott Hall, 1251 Wescoe Hall Dr, Lawrence, Kansas, 66045, USA}
\emailAdd{sergei@ku.edu}
\abstract{
A new numerical technique to identify the cosmic web is proposed. It is based on locating  multi-stream 
flows, \ie the places where the velocity field is multi-valued. The method is local in Eulerian space, simple 
and computaionally
efficient. This technique uses the velocities of particles and thus takes into account
the dynamical information. This  is in contrast with the majority of standard methods that use 
the coordinates of particles only. 
Two quantities are computed in every mesh cell: the mean and variance of the velocity field.
In the cells where the velocity is single-valued the variance must be equal to zero exactly, 
therefore the cells with non-zero variance are identified as multi-stream flows.  
The technique has been tested in a N-body simulation of the \L CDM model.
The preliminary analysis has shown that numerical noise does not pose a significant problem.
The web identified by the new method has been compared with
the web identified by the standard technique using only the particle coordinates.
The comparison has shown overall similarity of two webs as expected, 
however they by no means are identical. For example, the isocontours of the corresponding fields
have significantly different shapes and some density peaks of similar 
heights exhibit significant differences in the velocity variance and vice versa.
This suggest that the density and velocity variance have a significant degree of independence. 
The shape of the two-dimensional pdf of density and velocity variance confirms this proposition.
Thus, we conclude that the dynamical information probed by this technique introduces an additional 
dimension  into analysis of the web.
}


\begin{document}
\maketitle
\section{Introduction}
\label{sec:intro}
The morphology of the cosmic web is an active field of research in modern cosmology. 
A wide variety of methods have been suggested, see \eg
\cite{sh-etal-04,nov-etal-06,ara-etal-07,hah-etal-07,pla-etal-07,for-rom-etal-09,ara-etal-10,bon-etal-10a,bon-etal-10b}
just to mention a few. However, all these methods share one common feature: they study the cosmic web as
a property of the density field derived either from the distribution of  particles in space 
in cosmological N-body simulations or from galaxies in the redshift space. However, the web can be
viewed as a giant multi-stream flow. We suggest a simple technique that is capable to identify multi-stream flows and test it in a cosmological simulation of the \L CDM model.

We begin with  a short overview of the origin and evolution of the concept of multi-stream flows
in the context of the formation of the large-scale structure in the universe.
The notion of three-stream flows was introduced by Zeldovich \cite{z-70}, in his paper on pancakes.
He briefly discussed the formation of three-stream flows and illustrated the formation of a three-stream 
flow by the evolution of Eulerian coordinate as a function of Lagrangian coordinate with time.
However, he himself used it mainly as a useful auxiliary  theoretical tool helping 
to describe the nonlinear structures known at present as "Zeldovich's pancakes". 
Before 1980  he himself seemed not to think that  
the multi-stream flows as a physical phenomenon might have relevance to the formation 
of the structures in the universe. 
Similar views were common in the west as well. For instance, multi-stream
flows were barely mentioned in Peebles' book
published in 1980 and it was only in the context of critical remarks on the pancake model.

Doroshkevich et al. \cite{dor-etal-80} explored the evolution of the multi-stream flows in the one-dimensional 
N-body simulation a little beyond the three-stream flow stage up to seven streams. This simulation 
has showen that the multi-stream flow region remains relatively thin in the comoving 
coordinates in contrast with a simple extrapolation of the Zeldovich approximation (ZA) where it
grows unlimitedly.  The quantitative comparison of the 
thicknesses of the three-stream flow in ZA and in the 
one-dimensional numerical simulation was presented in the review \cite{sh-z-89}.
The geometry of generic caustics in ZA 
in two dimensions was discussed in \cite{a-sh-z-82}, the authors also provided a table of generic singularities 
occurring in the potential flows in three-dimensional space.
A detailed study of the phase space in self-similar gravitational collapse in 
one, two and three dimensions was presented in \cite{fil-gol-84}.  A high resolution two-dimensional
distribution of particles obtained in ZA that clearly showed the structure
of multi-stream flows was demonstrated in \cite{buc-89}, and in the high-resolution two-dimensional N-body simulations in \cite{mel-sh-89}.

The multi-stream flows were directly addressed in the model based on the adhesion approximation (AA)
 suggested in \cite{gur-etal-89}. 
The AA model was designed to control the runaway 
growth of the thickness of the pancakes in ZA
by introducing artificial viscosity term in the Euler equation. This term has no effect on
the motion in voids leaving it as it was in ZA but does not allow the 
formation of multi-stream flows by annihilating the transverse to pancakes or filaments velocities. 
Thus, the high density walls and filaments are formed in AA instead of multi-stream flows in ZA. 
The density profile controlled by the adopted value of the viscosity coefficient becomes smooth 
and the velocity remains single valued. It is worth stressing that  AA  
modifies the multi-stream flows by transforming them in single-stream flows of a special kind,
characterized by high density.
An important feature of the model consists in allowing the mass flow within the pancakes and filaments.
The model  incorporates most of the  features of the hierarchical clustering process which is
characteristic of the cosmological models dominated by cold dark matter.  
In addition, the AA model predicted several new features including the continuous flow
of mass from walls to filaments to clumps, the multiple merger of clumps, the collapse of some
voids, the presence of substructure of hierarchical type in voids, formation of the next generations
of the filaments and pancakes. It was quantitatively tested against the two-dimensional \citep{kof-etal-92} and three-dimensional \citep{mel-etal-94} N-body simulations and was also used for predicting 
the structure in the forthcoming SDSS \citep{wei-gun-90a,wei-gun-90b}.

Another significantly simpler modification of ZA  was suggested in \cite{col-etal-93}. 
Since the ZA model has a serious problem with the runaway growth of
the three-stream flow regions the authors proposed to filter out the perturbations with the
scales smaller than the scale of nonlinearity \ie the scale corresponding to the  r.m.s density 
contrast fluctuation being equal to unity. The tests of the model dubbed  the truncated 
Zeldovich approximation  (TZA) against three-dimensional N-body 
simulations generally confirmed what was expected: the large-scale structure was reproduced quite 
accurately but the structures on small scales were erased.  Two years earlier it had 
been shown that even replacing the small-scale perturbations with a new statistically independent 
realization would not change significantly the large-scale  structure \cite{lit-etal-91}. 
The large-scale structure
proved to be quite robust. The evolution of the structure with time could be crudely probed
by generating a sequence of particle distributions using  TZA
with the scale of nonlinearity corresponding to each chosen epoch. 

A very similar modification of ZA as far as the large-scale dynamics
is concerned  was suggested by Bond, Kofman and Pogosyan \cite{bon-etal-96} (the BKP model).  
The only but crucial  difference  between  the BKP and TZA models consists in  a different choice of the 
scale separating  the large-scale and small-scale dynamics. 
The authors  of  BKP model strongly stressed that the boundary
must be chosen in such a way that the large-scale motion was strictly  single-stream flow.
The authors  also emphasized the coherence of the filtered linear density field
and derived from it the strain tensor field on the scales of tens of Mpc between rare high density peaks, 
the idea somewhat similar to the  proposition made earlier in  \cite{dor-sh-78a,dor-sh-78b} 
and \cite{wei-gun-90b}.
The presence of this coherence causes the appearance of filaments and possibly pancakes.
However, the chosen condition for the scale separating large-scale dynamics from small scale dynamics
means that the filaments are the enhancements of density but not multi-stream flows. This is 
contrary to both the TZA and AA models.

It is probably worth mentioning one more difference between AA on the one hand and TZA and 
BKP on the other although it is  of a rather technical kind. The  TZA and BKP
models  similarly to ZA use the the strain tensor as a primary initial field 
(referred to as the deformation tensor in TZA).
The AA model is based on the linear velocity potential defined by the relation ${\bf v}= -\nabla  \Phi_v$,
 which coincides up to a
constant factor with the linear gravitational potential in the growing mode. 
In principle, both the velocity  potential and  strain fields contain exactly same information 
since one of them can be easily computed from the  other.
However, repackaging of the same information may be useful if it helps
to identify  the most significant variable that determines the structure or serves a particular
goal of the model better.
Summarizing this short discussion of three theoretical models of the large-scale structure
we stress one difference essential for this paper.  The multi-stream flows are characteristic for 
both  TZA and AA while the BKP model eludes them in the large-scale dynamics. 

The ZA model  defines  pancakes as the regions 
between the surfaces on which the density is formally singular \ie as the multi-stream flows.  
The three-dimensional N-body simulation by Klypin and Shandarin \cite{kly-sh-83} 
revealed that the most conspicuous features after clumps
are  filaments rather than pancakes.  Combining this finding with the results of \cite{a-sh-z-82} 
naturally caused the definition of filaments to become similar to that of pancakes: 
filaments are  the multi-stream flows having very oblong shapes. 
The similarity is based on dynamics, the both pancakes and filaments are nonlinear but unvirialized concentrations of mass therefore in the dynamical hierarchy they hold an intermediate position
between the density peaks that have not collapsed yet and virialized halos. 
At the same time another commonly used definition of objects as peaks above certain height 
in the density field  \citep{bar-etal-86} was also broadened to include pancakes and filaments 
\citep{bon-etal-96}. 
On the one hand the former is more physical but at the time seemed to be harder to implement 
especially when the dynamical information about the structure was scarce.
On the other the latter seems to be more practical but involves a free parameter, the threshold that 
determines which peak should be considered an object. 
This difference between two definitions was probably the reason of  the strongest 
discrepancy between  the models declared in \cite{bon-etal-96}: the formation of the objects
in BKS is in inverse order than in ZA.
It is worth pointing out that despite considerable overlapping
between two definitions  the objects they describe are not identical. 
 
Despite the  inability  of  current N-body simulations to resolve 
properly  the phase space there is no doubt that the multi-stream flows must be present in the 
collisionless medium in the nonlinear regime. 
Therefore even the limited information about multi-stream flows would introduce a new dimension
into the analysis of the cosmic web as it does in virialized halos \cite{mac-etal-10}. 
This must be based on using the velocity field. 
Velocities bring about dynamical information
totally independent of the density and gravitational potential fields in a general case.
There are of course  special cases
like the systems in virial or thermal equilibrium where certain relations of statistical
nature between velocities, coordinates of the particles and gravitational potential could be derived.

Recently there have already been made some attempts to incorporate dynamical 
information into the analysis of cosmic web  \citep{hah-etal-07, for-rom-etal-09}.
Both approaches are based on the analysis of the eigen values of the Hessian of the gravitational
potential at the nonlinear stage. Both groups appeal to the analogy with ZA
and claim that this statistics provides a dynamical classification of the cosmic web.
The only difference between two approaches is  in the different choices of the amplitude 
to be assigned to the eigen values of the Hessian.
However, as we have already mentioned neither density nor gravitational potential contain dynamic 
information in a general case since the both are computed from the particle 
coordinates only and absolutely  independent of  the velocities. 
We have already mentioned a couple of  special cases when the gravitational potential 
and velocities could be related in statistical sense.
The growing mode in the linear regime represents another special case.
The dynamical evolution  in ZA is
described by a map ${\bf x}({\bf q},t) = {\bf q} - D(t) \nabla\Phi_v ({\bf q})$,
where $\Phi_v$ is the velocity potential {\it not the gravitational potential}. 
It is only due to the degeneracy of the growing 
mode in the linear regime where the velocity potential proportional
to the gravitational potential  $\Phi_v = constant \times  \Phi_g$ they can be used
interchangeably with the proper choice of the constant.
Therefore the gradient of the gravitational potential is proportional to the velocity
and  it can be computed from the gravitational potential in the growing mode
in the linear regime. The ZA model is of course an extrapolation
and therefore required additional scrutiny before it was excepted as a reasonable
approximation for the initial phase of the nonlinear regime (see  \eg \cite{sh-z-89}).
However, neither the gradient of the potential can be used for mapping
nor  its second derivatives represent the deformation tensor in a general case.
If correct the reasoning presented in \cite{hah-etal-07} and \cite{for-rom-etal-09} must be valid for
arbitrary velocities of the particles since the analysis and thus the conclusions do not depend 
on the velocities at all.  This hardly can be true. 
Thus, even the suggested statistics may provide some merits 
for the study of the density and gravitational potential fields at the nonlinear stages 
its interpretation as a dynamical characteristics cannot be excepted in the present form.

The goal of the paper is to introduce a new numerical technique to identify 
the cosmic web as the multi-stream flows and present the results of the first tests, which are
very encouraging. The method is based on a simple statistics derived from the coordinates 
and  velocities of the particles: the mean and variance of velocities in every mesh cell.
Here we use a uniform spatial mesh although the homogeneity of the mesh is not required by the method.
We estimate the role of noise caused by numerical effects and
compute the two-dimensional pdf of the density and variance of velocity
which demonstrates a significant level of independence of these quantities.
We compare the appearance of the multi-stream flow web 
with the web  derived from the density field only as well as with the distribution of the particles themselves.

Although there is no agreement between different groups on the exact definition of the filaments
and walls/panckes
the most agree that the filaments comprise more mass than any of the rest constituents of the
web (\ie clusters/clumps, walls/pancakes, field/void). For instance,  the fraction
of mass in filaments and walls in the \L CDM simulations to be almost  50\% of the total mass, 
considerably more than in clusters \cite{ara-etal-07}.   Both filaments and walls are nonlinear but 
unvirialized objects in contrast to clusters/clumps which are mostly virialized objects. 
The suggested method allows to
quantitatively analyze the dynamics of these objects because they are  multi-stream flows.
This work is under way and the results  will be reported in detail in a separate paper.

The rest of the paper is organized as follows. Section \ref{sec:1Dexample} describes the
idea of the method, Sec. \ref{sec:numer-tech} describes the numerical technique and 
N-body simulation used in the tests, the noise analysis is given in Sec.  \ref{sec:noise}.
The total volume devoid of multi-stream flows is evaluated in 
Sec. \ref{sec:void-threshold}. The appearance of the multi-flow web is compared 
with the density based web in Sec.  \ref{sec:pictures}, and the mean velocity field is shown in
Sec. \ref{sec:mean-vel}. Finally the results are summarized in Sec. \ref{sec:summary}.
In the rest of the paper density is given in the units of the mean dark matter density, velocities in
km/s and \sv in (km/s)$^2$.

\section{ The idea of the method}
\label{sec:1Dexample}
\begin{figure}
\centering\includegraphics[width=17cm]{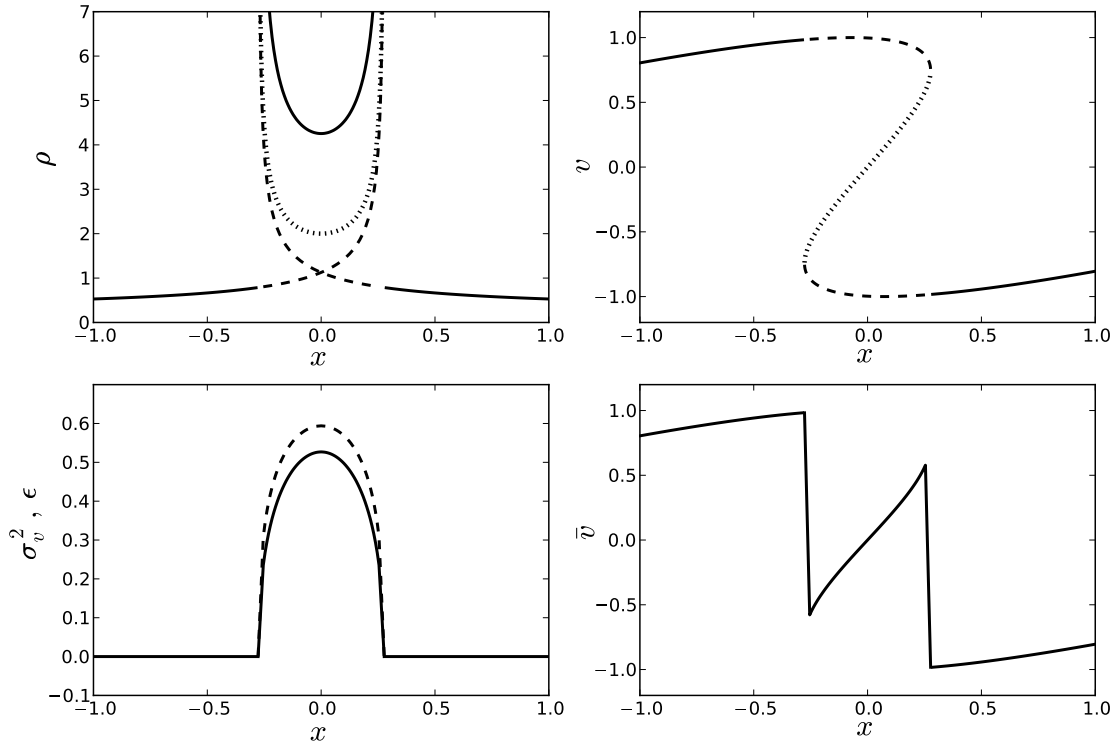} 
\caption{Top left panel. The density as a function of the Eulerian coordinate, $x$, is shown 
for all  streams by dotted and dashed lines. The total density is shown by a solid line.
Top right panel. The velocity in three streams are shown by dashed and dotted lines,
the solid line shows the velocity outside the three-stream flow.
Bottom left panel. The solid and dashed lines show respectively  $\sigma_v^2$ and $\epsilon$ defined in eq. \ref{eq:sigma}. Bottom right panel shows the mean velocity defined in 
eq. \ref{eq:mean-vel}. 
}
\label{fig:pancake}
\end{figure}

The physical idea of the method can be illustrated by a simple example.
Let us consider the formation of a one-dimensional pancake in ZA 
for a simple sinusoidal perturbation \citep{z-70,a-sh-z-82,sh-z-89}. 
Using the comoving coordinates
one can easily write down explicit expressions for  the position, velocity and density as a function of 
the Lagrangian coordinate $q$ and   the linear growth factor $D$ that can effectively play the role of time 
\b
x = q - D \sin q, ~~~ v \equiv dx/dD = -\sin q, ~~~ \rho = \left| 1-D \cos q \right| ^{-1}.
\e
The solution predicts a shell-crossing singularity in density at $x = 0$ at $D=1$ and the formation of 
a three-stream  flow afterward. At small $\delta D = D - 1$ it approximately describes the generic phase space structure of the shell crossing regions at their early stages. At later times it evolves into
a multi-stream flow with five then seven etc. streams which is beyond the scope of ZA. 
The density and velocity profiles in the three-stream flow and its
vicinity are shown in the top panels  in Fig. \ref{fig:pancake}. 
The velocity and density are shown for each stream in addition 
the total density in the three-stream flow region is also shown.
Two bottom panels show the mean velocity, velocity variance $\sigma ^2$ (solid line) and the density of kinetic energy  $\epsilon$ (dashed line) defined bellow.

Usually the analysis of the structure of multi-stream flows is limited to the discussion 
of velocity and density fields. 
We propose to examine  two additional quantities of thermodynamic and therefore statistical kind
similar to density.
One of them is the variance of velocity, $\sigma_v^2(x)$, and the other is the mean kinetic energy $\epsilon(x)$ at every Eulerian point $x$
\b
\sigma ^2_v(x) \equiv {\sum_{i=1}^{n_s} \rho_i(x) \Delta v^2_i(x)\over \sum_{i=1}^{n_s} \rho_i(x)}, ~~
\epsilon(x) \equiv {\sum_{i=1}^{n_s} \rho^2_i(x) \Delta v^2_i(x)\over \sum_{i=1}^{n_s} \rho_i(x)},
\label{eq:sigma}
\e
where $\rho_i(x)$ and $v_i(x)$ are the density and velocity in each stream and 
$\Delta v_i \equiv v_i(x) - \bar{v}(x)$, $n_s$ is the number of streams. The mean velocity is defined as
\b
\bar{v}(x) \equiv  {\sum_{i=1}^{n_s} \rho_i(x) v_i(x)\over \sum_{i=1}^{n_s} \rho_i(x)}.
\label{eq:mean-vel}
\e
Figure \ref{fig:pancake} shows all these quantities at the nonlinear
stage when the shell crossing has already happened and the three-stream flow  has formed.
Passing by we note that
the mean velocity of the three-stream flow does not coincide with the velocity in one of the streams
as may appear in the figure, but they are quite similar in this particular example.

Comparing the shapes of $\sigma ^2_v$ and $\epsilon$ curves with the density and velocity curves we
see remarkable  differences. 
The most important, however not  surprising feature of the  
$\sigma ^2_v$ and $\epsilon$ curves is that they are identical zeros beyond the multi-stream flow
regions. Less obvious and more surprising feature is that their shapes are so similar to each other and practically inverse to the shape of the total density curve in the shell crossing region. The both features
suggest that the $\sigma ^2_v$ and $\epsilon$ functions could be complimentary characteristics
to the density in the studies of the dark matter cosmic web.
The fact that  they are equal exactly to zero beyond the multi-stream flow regions 
may mean that they can be good tracers of multi-stream flows.
However, the discussed example is too oversimplified.  In the reality the noise caused by discreteness
undoubtedly  violates this ideal situation and therefore must be taken into account before 
one can come to a sound conclusion.  But first we describe the numerical technique.

\section{The numerical technique and N-body simulation}
\label{sec:numer-tech}
\subsection{Numerical technique}
Since $\sigma ^2_v$ and $\epsilon$ are so similar we shall limit the discussion to $\sigma ^2_v$
only. 
The velocity variance depends on density less than $\epsilon$ and therefore it carries more  independent 
information. The physical dimensions of \sv suggest that it can be directly related to the gravitational 
potential. However a more thorough analysis may reveal some additional attractive features
of $\epsilon$.

We use the standard CIC (cloud-in-cell) method for the evaluation of $\bar{v}$ and then
$\sigma ^2_v$.  The  particles are modeled as constant density cubes of size $l=L/N_p$ 
where $L=$512 \hm1 is the size of the simulation box and $N_p=$512 is the number of particles in one dimension. Both $\bar{v}$ and
$\sigma ^2_v$ are evaluated on a uniform cubic mesh with the cells of the same size as that of particles.
However, it is worth mentioning that the uniformity of the mesh is not a requirement of the method.
In general each particle contributes a volume weighted fraction of its velocity to eight neighboring
 mesh sites.
Each velocity fraction equals the fraction of the volume of the overlap of the particle cloud  
with the mesh cell.
It also can be viewed as a fraction of its linear momentum since all the particles have the same masses. 
The mean velocity assigned to the mesh point is the sum of the contributions from all the particles
overlapping with it divided by the total mass in the cell. The variance $\sigma_v^2$ is evaluated in a 
similar manner. It is worth stressing that we conduct the further analysis without additional
filtering of the density field which is  also computed using the CIC scheme.
\subsection{The cosmological model and simulation parameters}
We applied this method to the pure dark matter N-body simulation in the 512 \hm1 Mpc cubic
box using PM (particle mesh) code \citep{hab-etal-09,pop-etal-10}. The number of particles was 512$^3$ and the mesh in the gravitational force solver was 1024$^3$. The parameters of the \L CDM cosmological
model were as follows: $h = H_0/(100 \, km/s\cdot Mpc) = 0.72$, $\Omega_{tot} = 0.25$, $\Omega_b= 0.043$,
$n = 0.97$, $\sigma_8 = 0.8$, the initial redshift $z_{\rm in} = 200$. The choice of the parameters is related  to the main purpose of the study:
to test this new technique and illustrate its performance by applying it to a realistic cosmological model.

\section{The analysis of numerical noise}
\label{sec:noise}
We begin with applying this method to the initial state when all the fields are in the
linear regime.
\subsection{Initial linear state}
\begin{figure}
\begin{minipage}[t]{.99\linewidth}
  \centering\includegraphics[width=7.5cm]{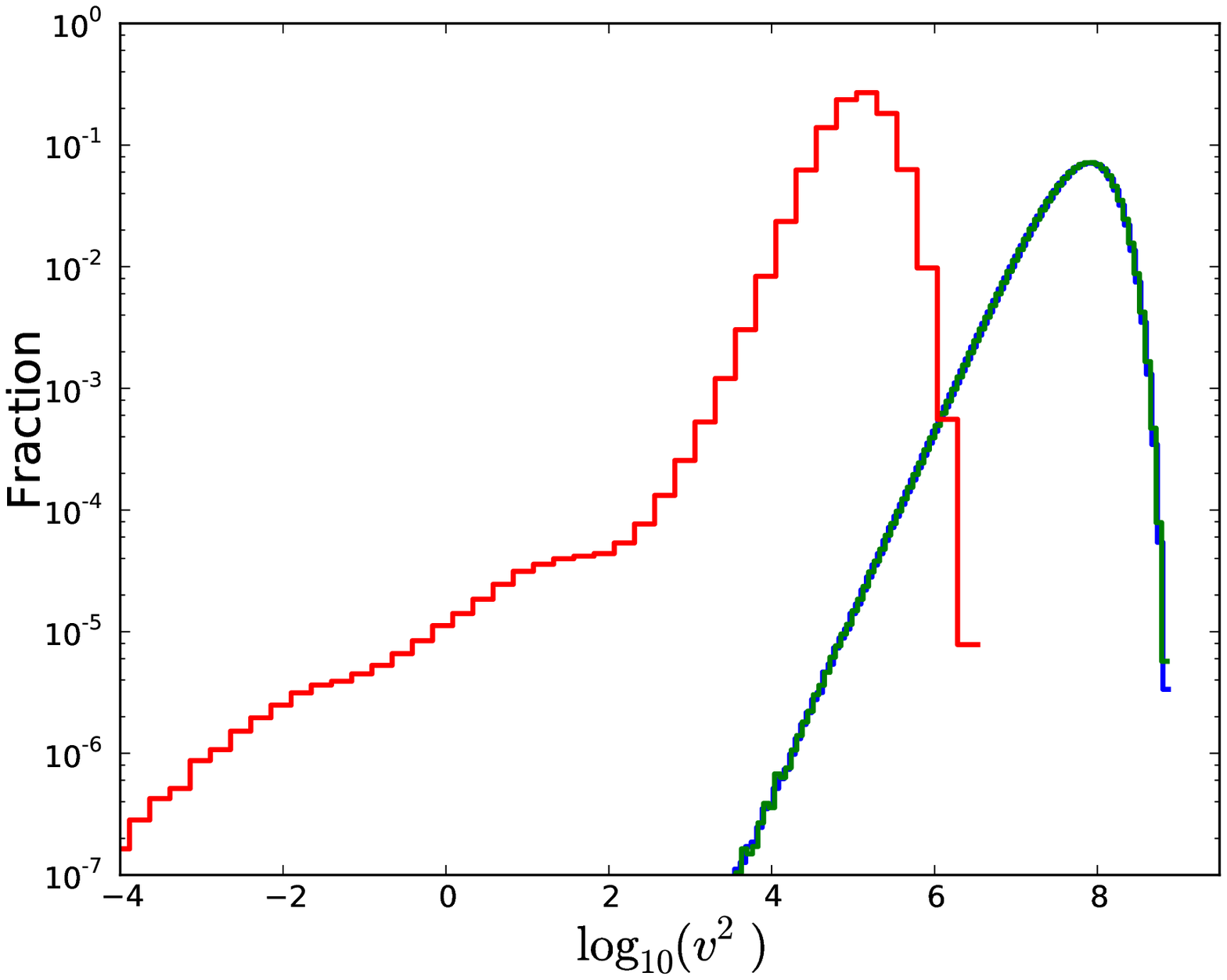} 
  \includegraphics[width=7.5cm]{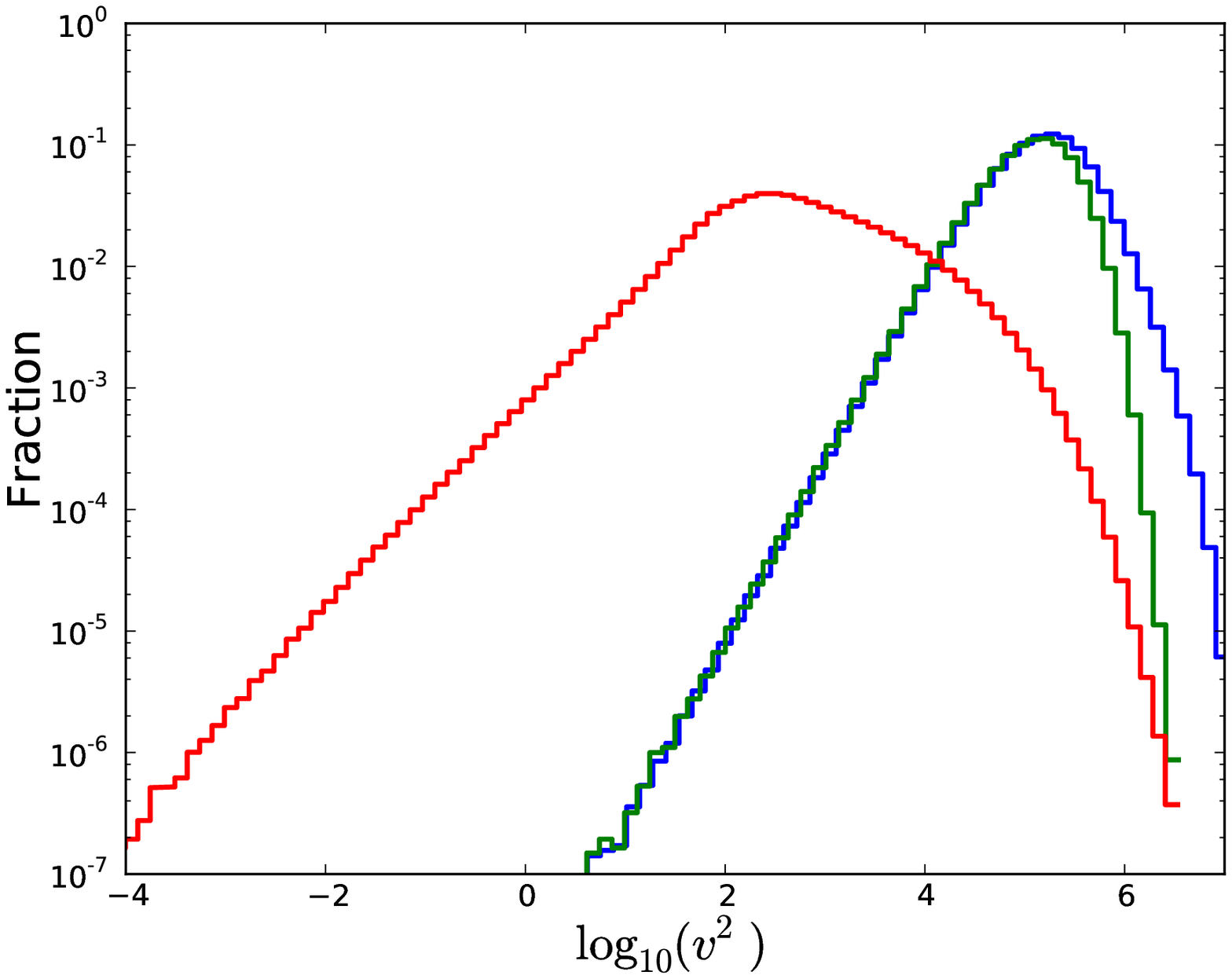}
\end{minipage}\hfill
\caption{The fraction of particles per bin as a function of $v_p^2$  (blue histograms), 
the fraction of mesh cells per bin as a function of  $\bar{v}^2$ (green histograms), and as a function of
$\sigma_v^2$ (red histograms) are shown for the initial state in the left hand side panel and for the present time in the right hand side  panel. The fraction of particles with $v_p^2$ and the fraction of cells with $\bar{v}^2$ since $v_p^2 = \bar{v}^2$ in the one-stream flow regions, thus the green and blue curves are one on top of the other in the left panel.  There are  also 33.4\% of cells with $\sigma_v^2 < 10^{-9}$ not shown
 in the right hand panel. 
  }
\label{fig:hist}
\end{figure}

Even at the very beginning the small displacements of particles from unperturbed positions on
the regular grid result in small overlapping of particles with each other. It is also easy to
imagine that more than one particle may overlap a mesh cell. This results in generating of
small but nonetheless nonzero $\sigma_v^2$.  The left hand side panel in Fig. \ref{fig:hist}
shows three functions $f(v_p^2)= n(v_p^2)/512^3$ (in blue), $f(\bar{v}^2)$ (in green) and $f(\sigma_v^2)$ (in red),
which are the fractions  of particles or cells  per equally sized logarithmic bins. In the ideal situation
one expects that $f(v_p^2) = f(\bar{v}^2)$ and $f(\sigma_v^2) \equiv 0$ as it was described in
Sec. \ref{sec:1Dexample}. The first condition is satisfied with high accuracy and the second
is obviously violated. However, the $\sigma_v^2$ is approximately three orders of magnitude
smaller than $v_p^2$ or $\bar{v}^2$ therefore we conclude that this test is passed without serious
problem. In the multi-stream flows the \sv is expected to be less than $v_p^2$ but not orders of
magnitude smaller (see Fig. \ref{fig:pancake} for an illustration). 

\begin{figure}
\begin{minipage}[t]{.99\linewidth}
  \centering\includegraphics[width=7.5cm]{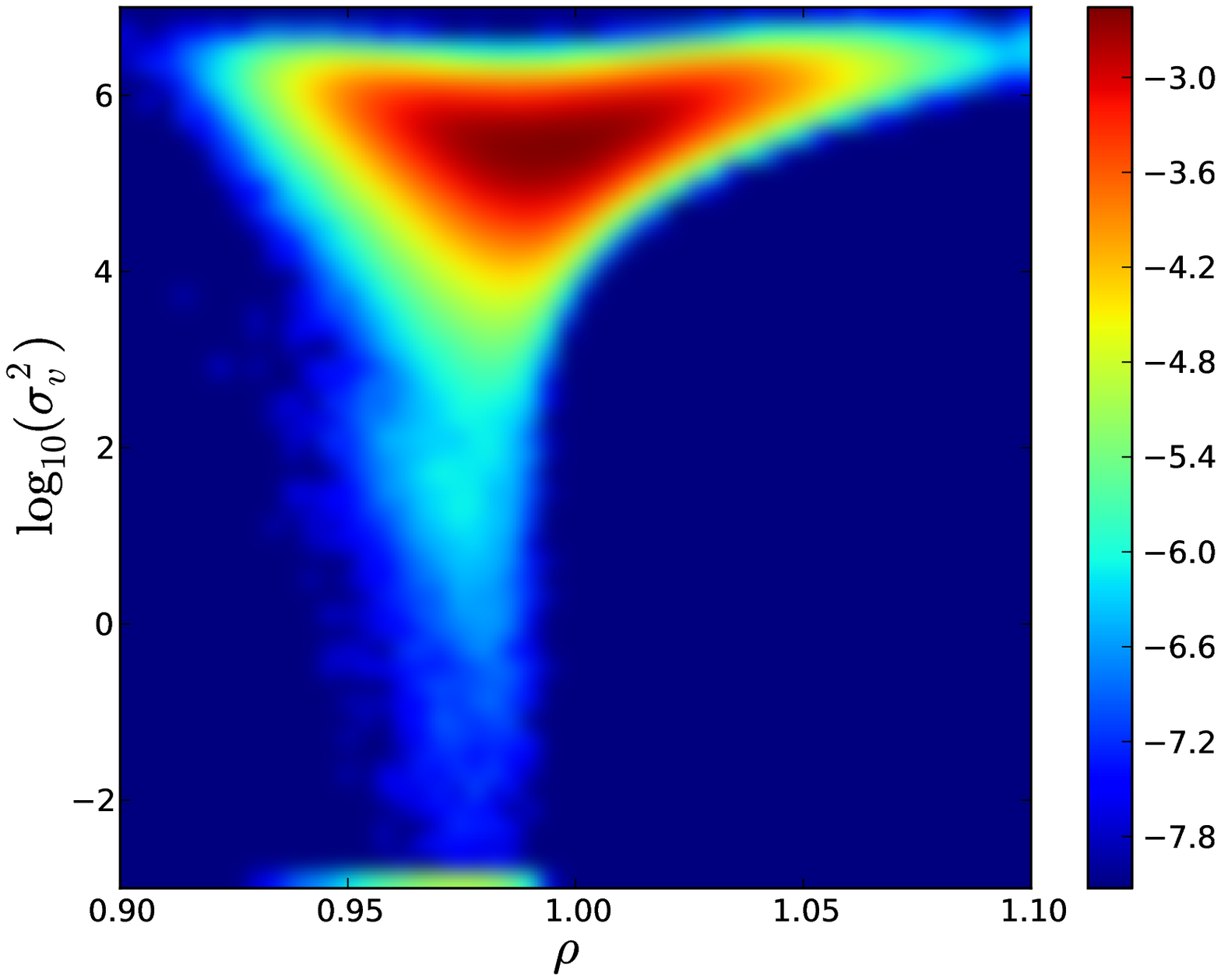} 
  \includegraphics[width=7.5cm]{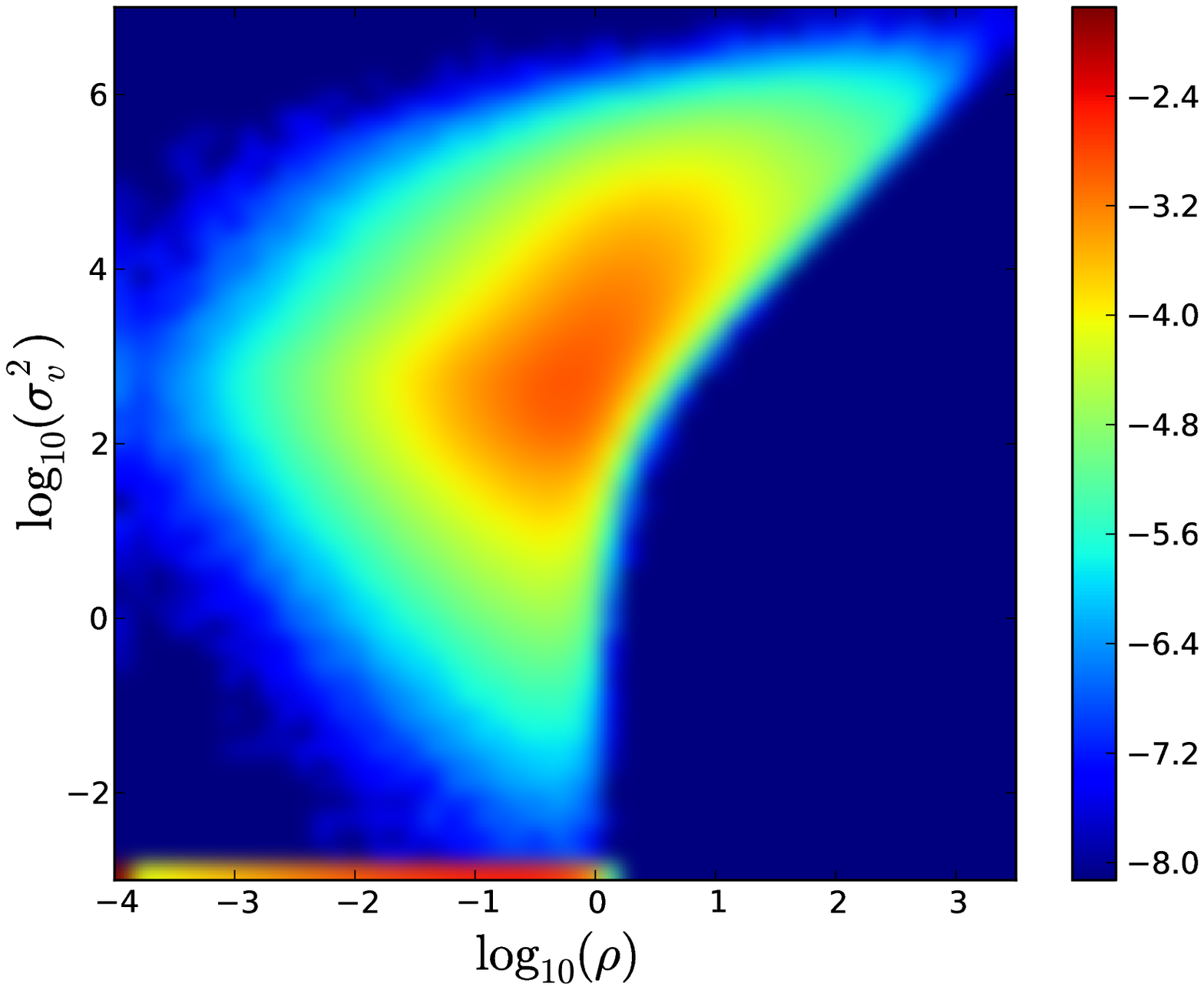}
\end{minipage}\hfill
\caption{The fraction of cells $n(\rho, \sigma_v^2)/512^3$ in bins as a function of $\rho$ and $\sigma_v^2$. 
The color bars show the logarithmic scales for each panel. The both histograms are of same size 50$\times$50 with equally spaced bins along the axis marked in each panel.
Left:  the initial linear state. Right: the final highly nonlinear state.
}
\label{fig:hist2d}
\end{figure}

An additional question may occur. Does the generation of noise in $\sigma_v^2$ depend on the
density in the cell? Naively one may think that \sv is generated in the regions where particles
are crowded and much less in the underdense regions with $\rho <1$ where the particles are
deserted.
The answer to this question happenes to be not that simple as the left hand side panel in Fig. \ref{fig:hist2d} shows.
The figure displays the two-dimensional histogram $f(\rho, \sigma_v^2)$  for the initial state. 
The maximum of the function is at
$\rho \approx1$, $\log_{10}(\sigma_v^2) \approx 5.5$. The peak has approximately a triangle
shape, extending to somewhat higher values of $\sigma_v^2$ for both higher and lower
densities. The third direction is downward to lower values of $\log_{10}(\sigma_v^2)$. 
This relatively complicated
shape of  $f(\rho, \sigma_v^2)$ indicates that the kinematics of particles is not simply contraction
for $\rho >1$ and expansion for $\rho <1$ even at this early stage. The overall shape of the histogram
shows no strong correlation between $\sigma_v^2$ and $\rho$.
\subsection{Final nonlinear state}
The statistics  $f(v_p^2)$,  $f(\bar{v}^2)$, $f(\sigma_v^2)$, and $f(\rho, \sigma_v^2)$ 
look quite differently at the nonlinear stage. The right hand side panel in Fig. \ref{fig:hist} 
shows the histograms
for the first three functions. Only  $f(v_p^2)$ and  $f(\bar{v}^2)$ bellow the maximum are similar. 
At the high value end $f(\bar{v}^2)$ drops considerably faster than $f(v_p^2)$ since the mean velocity
in the multi-stream flow regions is significantly lower then the velocities of particles
(see also  Fig. \ref{fig:pancake}). The high value end of $f(\sigma_v^2)$ is much closer to that of both 
$f(v_p^2)$ and $f(\bar{v}^2)$ as one may naturally expect for the multi-stream flow regions.
It is worth mentioning that the highest mean velocities are greater than the highest relative
velocities inside multi-stream flows. 

As many as 11.76\% of cells have $\sigma_v^2 = 0$ exactly in excellent agreement with 
the fraction of completely empty cells with $\rho = 0$ being equal to 11.74\% , 
in addition there are 21.6\% of cells with $0< \sigma_v^2 <  1 \times 10^{-9}$. 
This explains the presence of red bar at the bottom of Fig. \ref{fig:hist2d} (right panel).
The two-dimansional histogram $f(\rho, \sigma_v^2)$ shown in the right hand side panel 
of Fig. \ref{fig:hist2d} is also substantially different from the linear case. There is an easilly
seen correlation between $\sigma_v^2$ and $\rho$, at $\rho > 1$.  However, the spread 
is about two orders of magnitude along the both directions for the red peak. The former is naturally
related to the correlation of the gravitational potential depth with the high density peaks. The latter
suggests that many medium density peaks ($ 10 < \rho < {\rm (a ~ few)} \times 100$) as well as the
filaments are not relaxed systems with \sv in such  cells to be almost independent of $\rho$.
\section{The total volume in voids}
\label{sec:void-threshold}
\begin{figure}
\centering\includegraphics[width=10.cm]{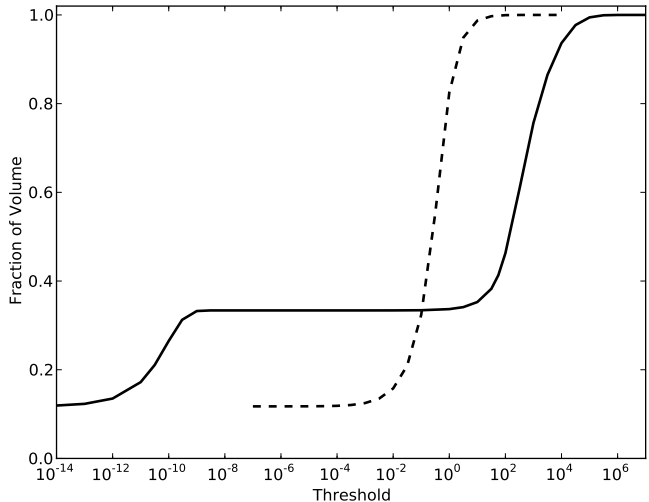} 
\caption{The fraction of volume where $\sigma_v^2$ (solid line) or $\rho / \bar{\rho}$ (dashed line) is less
 than the threshold. The horizontal gives the threshold in (km/s)$^2$ for  $\sigma_v^2$  and in dimensionless
 units for $\rho/\bar{\rho}$. The low limit for both quantities is the same 11.7\% 
 which is the fraction of sites where the density is zero exactly in the N-body simulation. 
}
\label{fig:void_threshold}
\end{figure}
There are 11.74\% of the mesh cells completely empty, they obviously have $\sigma_v^2=0$ exactly
as well (except a tiny fraction due to numerical errors). 
However, additional  21.6\% of cells have such a tiny $\sigma_v^2 < 10^{-9}$  that they can
be considered belonging to one-stream flow regions without doubts. 
The fractions of cells one as a function of density and  the other as a function of $\sigma_v^2$ 
where the corresponding quantity is less then the threshold shown in the horizontal is shown  
in Fig. \ref{fig:void_threshold}.
The function of  $\sigma_v^2$ grows from 33.24\% at $\sigma_v^2 = 1\times10^{-9}$ 
to 33.67\% at  $\sigma_v^2 = 1$ and to 35.26\% at $\sigma_v^2 = 10$. 
Then the rate of growth becomes significantly greater and the fraction $f(\sigma_v^2)$  
reaches 46.27\% and 60.70\%  at $\sigma_v^2 = $100 and 316 respectively. 
The long plateau from $\sigma_v^2 = 10^{-9}$ to
about 10  (ten orders of magnitude!) shows that in about one third of the volume the velocity is a single valued function, \ie no multi-stream flow regions are present there. This number can serve as a low 
physical limit for the total volume in voids for the resolution of this simulation. 
For a given mass resolution of N-body simulations one can give a physical definition of voids as the regions where the shell-crossing has not yet occurred. 
Therefore neither halos nor filaments nor sheets could have formed in these regions. 
The multi-stream flow regions can apparently appear in such places with the growth of the resolution
of the simulation.
\begin{figure}
\centering\includegraphics[width=18cm]{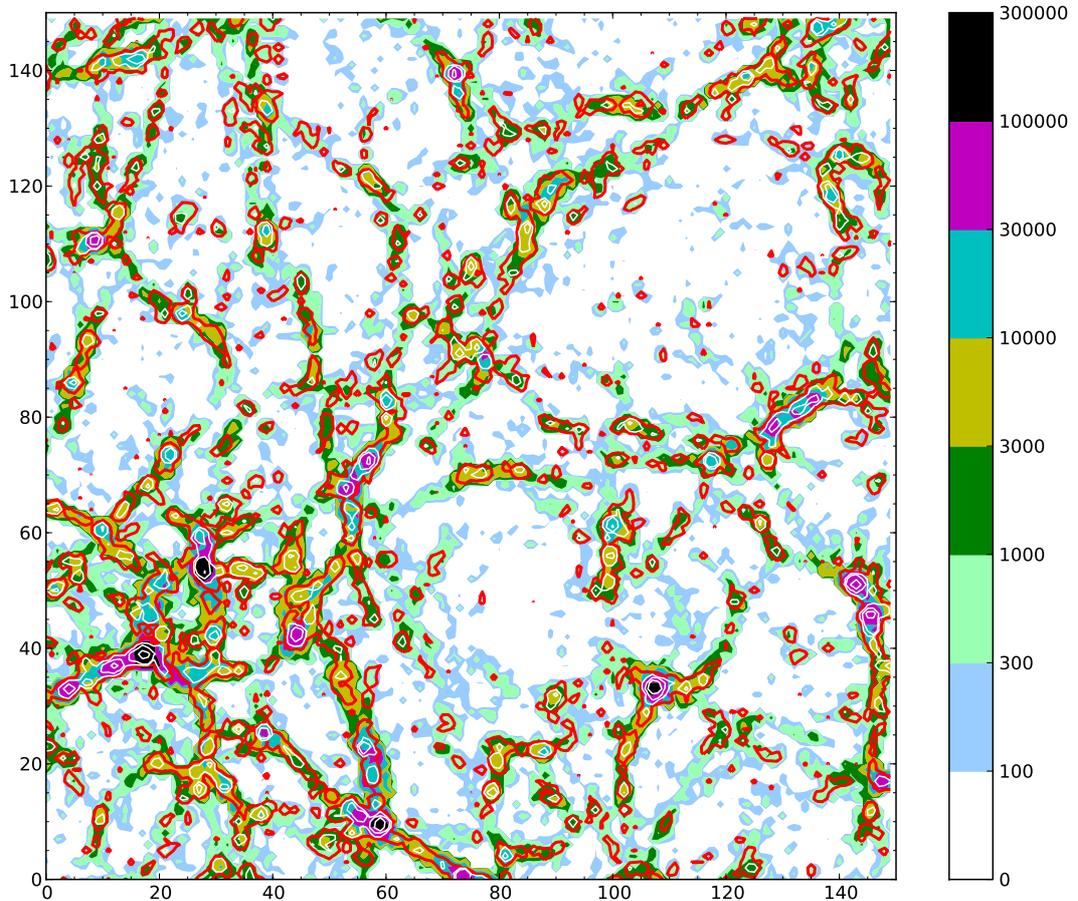} 
\caption{A part of a thin slice of 1 $h^{-1}$ Mpc thickness through N-body simulation box. 
The filled contours show the values of $\sigma_v^2$ as indicated by the color column.
Heavy red contours  show the CIC density field with $\rho = 1$, the white thin contours 
inside red contours show $\rho= 3, 10$  and $50$
The labels show the distances in $h^{-1}$ Mpc.
}
\label{fig:slice150}
\end{figure}
\begin{figure}
\centering\includegraphics[width=18cm]{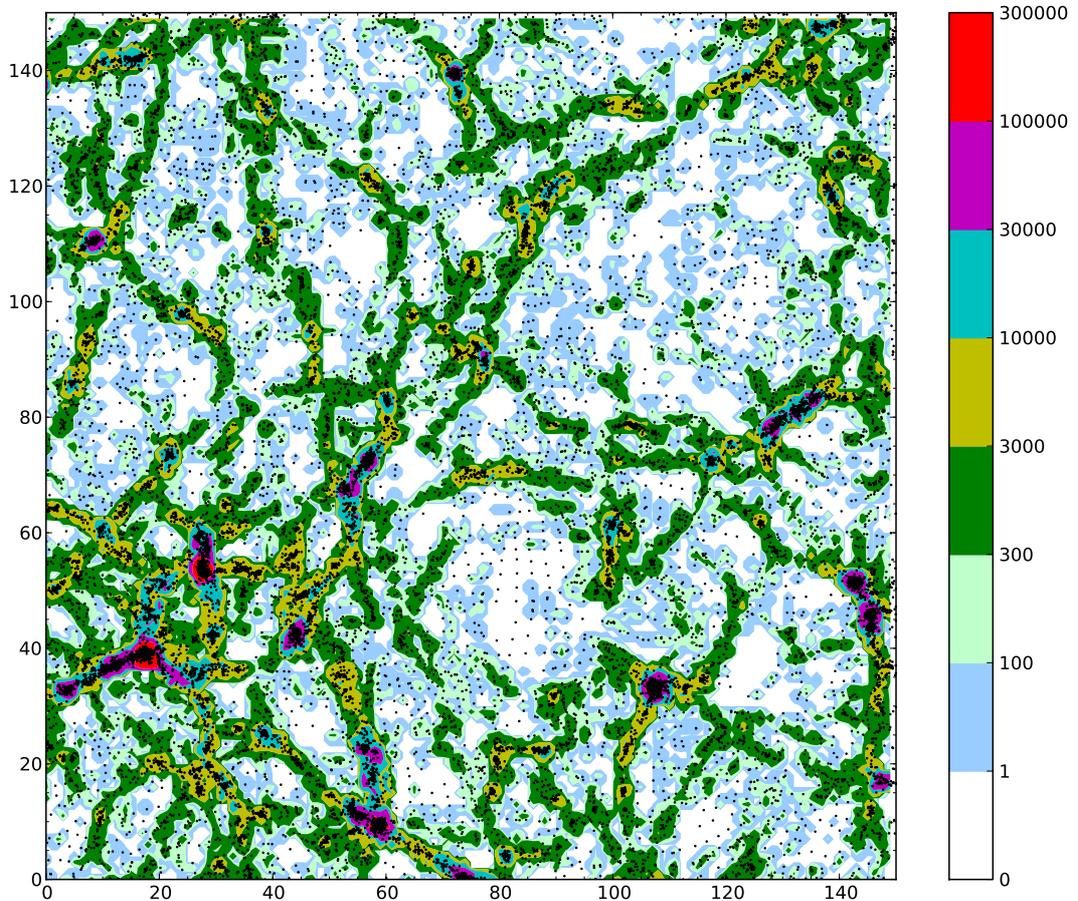} 
\caption{The same slice as in Fig. \ref{fig:slice150}. 
The filled contours show the regions with $\sigma_v^2$ displayed by the color column.
 Dots are the particles contributed to the density and $\sigma_v^2$ in this slice.
}
\label{fig:part_slice150}
\end{figure}
\section{The multi-stream flows  versus the density web}
\label{sec:pictures}
In this section we compare the multi-flow web with the standard representation of the web by the
isocontours  of the dark matter density.
Figure \ref{fig:slice150} shows the density contours superimposed on the field of $\sigma_v^2$ in a 
thin slice (1 \hm1 Mpc) randomly selected from the simulation cube. The filled contours show the
regions of the $\sigma_v^2$ field between the levels marked on the color column. The  density contours
correspond to the following levels: red  to $\rho = 1$ and three white contours correspond to $\rho = 3, 10$
 and $50$. One can see that although the overall structures are very
similar, but they by no means are identical. We consider the both observations as good news. 
The former is the evidence of the robustness of the identification of the filaments and other structures 
while the latter suggests that dynamical information in the \sv field brings an additional dimension
to the characteristics of the web. 
A closer inspection show that the highest peaks in two fields do not have identical shapes. 
It will be better seen in the following figures where
we zoom up two regions: one with the lowest mean density (void region) and the other with the highest density in the slice (clumps and filaments).

However, we first show the superposition of particles contributing to the density and $\sigma_v^2$
in this slice on the  $\sigma_v^2$ field in Fig. \ref{fig:part_slice150}. The color map is slightly
different from the previous figure, which allows to see the structures at low values of $\sigma_v^2$
better. In general, we conclude that the correspondence of two fields it very good as it should be
apart from numerical noise.
\begin{figure}
  \centering\includegraphics[width=17.5cm]{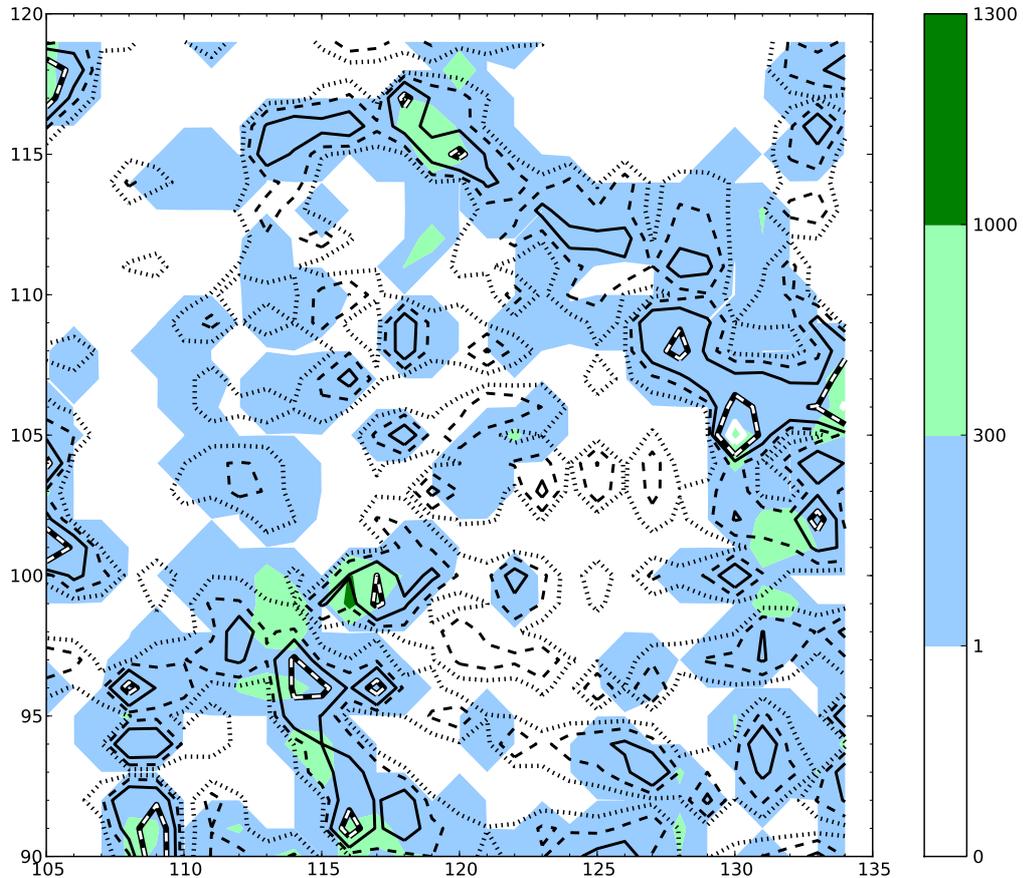} 
\caption{A zoomed low density region in the top right corner of  Fig. \ref{fig:slice150} 
and \ref{fig:part_slice150}. 
The filled contours show the values of $\sigma_v^2 $ as indicated by the color column.
The black and white contour lines show a few density peaks above unity in units of the mean density.
 The solid, dashed and dotted lines  correspond to $\rho =$ 0.5,  0.25 and  0.1 respectively.
 The heavy white line at about (130, 105) shows one small contour  with  $\rho = 2$. 
}
\label{fig:void30}
\end{figure}
\subsection{The slice through a void}
The region from the top right corner of the slice is zoomed in Fig. \ref{fig:void30}.
The color map is changed again in order to adopt better for the low density environment. The levels of
$\sigma_v^2$ are shown by the color column while the density contours  are shown by black lines. 
The density in the most of the region is below the mean except a few peaks shown by black and white 
heavy line. The impression is similar to that from Fig. \ref{fig:slice150}: the fields have many similarities but also substantial differences, perhaps even a little greater differences than in Fig. \ref{fig:slice150}.
In particular, the empty regions seem to be significantly more empty in $\sigma_v^2$ than in $\rho$
in an agreement with the discussion in Sec. \ref{sec:void-threshold}.
Although the density contours are pushed to such a low values that one may completely disregard them
as being subgrid noise the values of the $\sigma_v^2$ contours are also very low and still there is
some similarity between both fields in some parts of the figure. Thus, an optimistic conclusion would
be that the combination of both fields may allow to probe low density regions more reliably than any of them alone. On the other hand, a pessimistic conclusion would relate it to the correlation of the $\rho$ and 
$\sigma_v^2$ fields at small $\sigma_v^2$. The further study is obviously required. 
\begin{figure}
  \centering\includegraphics[width=17.5cm]{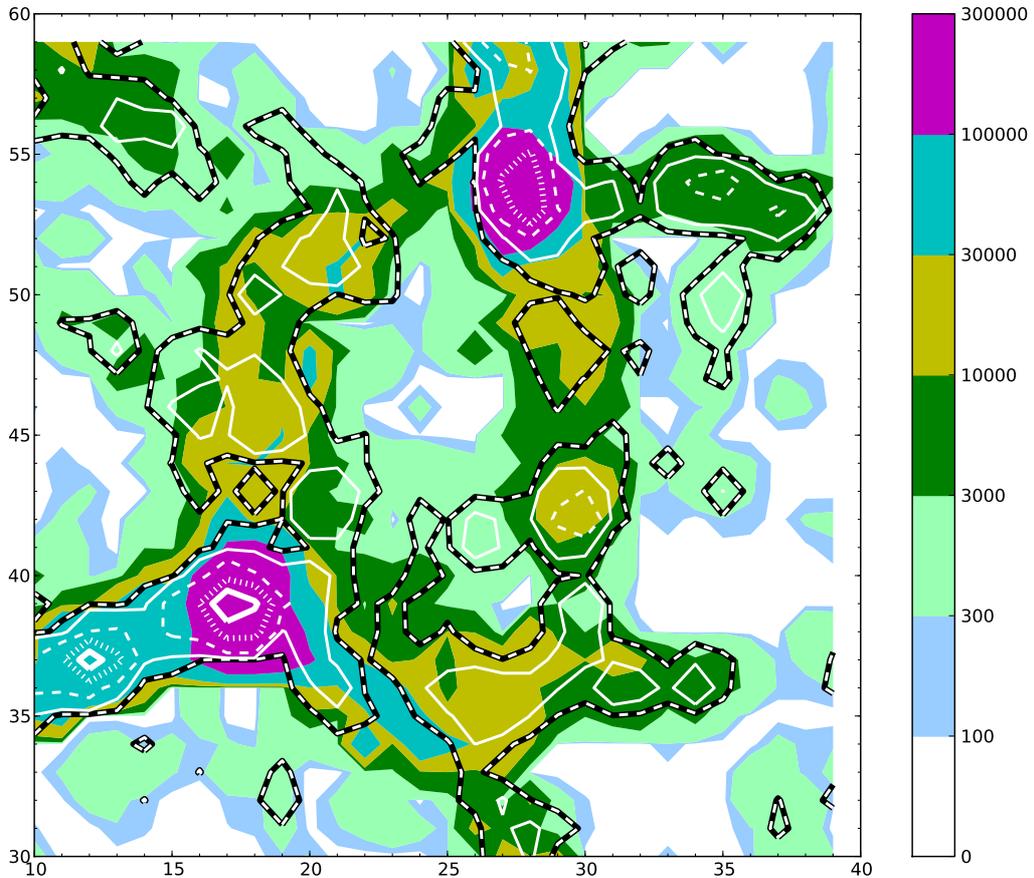}
\caption{A zoomed high density region in the bottom left corner of  Fig. \ref{fig:slice150}
and \ref{fig:part_slice150}.
Here, filled  contours  correspond to $\sigma_v^2$ as indicated by the color. The black and white line 
shows contours of $\rho$ = 1, and white solid, dashed, dotted and heavy solid  contours 
correspond to  $\rho=$3 , 10 , 30 , and 60 respectively.
}
\label{fig:clumps30}
\end{figure}

\subsection{The slice through clumps and filaments}
Figure \ref{fig:clumps30} shows the relatively high density region in the bottom left corner of Fig. \ref{fig:slice150} and \ref{fig:part_slice150}.  The figure contains a few density peaks with $\rho >10$ 
(white dashed lines) three of which are higher than $\rho = 30$ and two even higher with $\rho > 60$.
There are also a few filaments connecting the clumps. Again the both fields are rather similar but there
are notable differences as well. Two highest density peaks in the bottom left corner reach $\rho > 60$ but only 
one exceeds $\sigma_v^2 > 10^5$. However, the peak of $\sigma_v^2$ extends to the density level 
as low as the mean density. On the other hand a little lower density peak with $\rho < 60$ coincides 
with a quite substantial region with $\sigma_v^2 >10^5$. Similar observations can be made with the peaks
of lower density. Thus, we conclude that the density and $\sigma_v^2$ are two parameters that 
have a substantial  degree of independence and therefore may better characterize the clumps and
filaments as any of them alone. In addition, the figure shows that the filaments are the multi-stream flows
in agreement with both the TZA and AA models and contrary to the prediction of the BKP model
that implies that the filaments have not reached the state of multi-stream flows.

\section{The mean velocity field}
\label{sec:mean-vel}
Finally, we present the mean velocity field defined by eq. \ref{eq:mean-vel}. 
We superimpose the mean velocity field shown by the arrows on Fig. \ref{fig:clumps30} 
where we keep only three highest levels of the density contours for clarity. 

The highest density peak  sitting at the largest patch of high \sv is in the bottom left corner
approximately at (17, 39).
The overall pattern of the mean velocity  field clearly indicates that the whole region shown in the
figure is falling on this clump. The mass inside three filaments is streaming toward the clump
confirming one of the predictions of the AA model \citep{gur-etal-89}.
The second largest clump at (28, 54) is falling on the first clump accompanied  by the surrounding filaments.
The filaments are highly inhomogeneous and the streams are far from being uniform. 
The velocity field in the  vicinity of the first clump looks roughly circular giving
the impression of quasi-spherical (actually quasi-circular in this plane because we see only 
the cross section) accretion but this is misleading if we look at the flux of mass  $\rho \bar{\bf v}$
shown by arrows in Fig. \ref{fig:mass-flow-clumps30}.
\begin{figure}
  \centering\includegraphics[width=17.5cm]{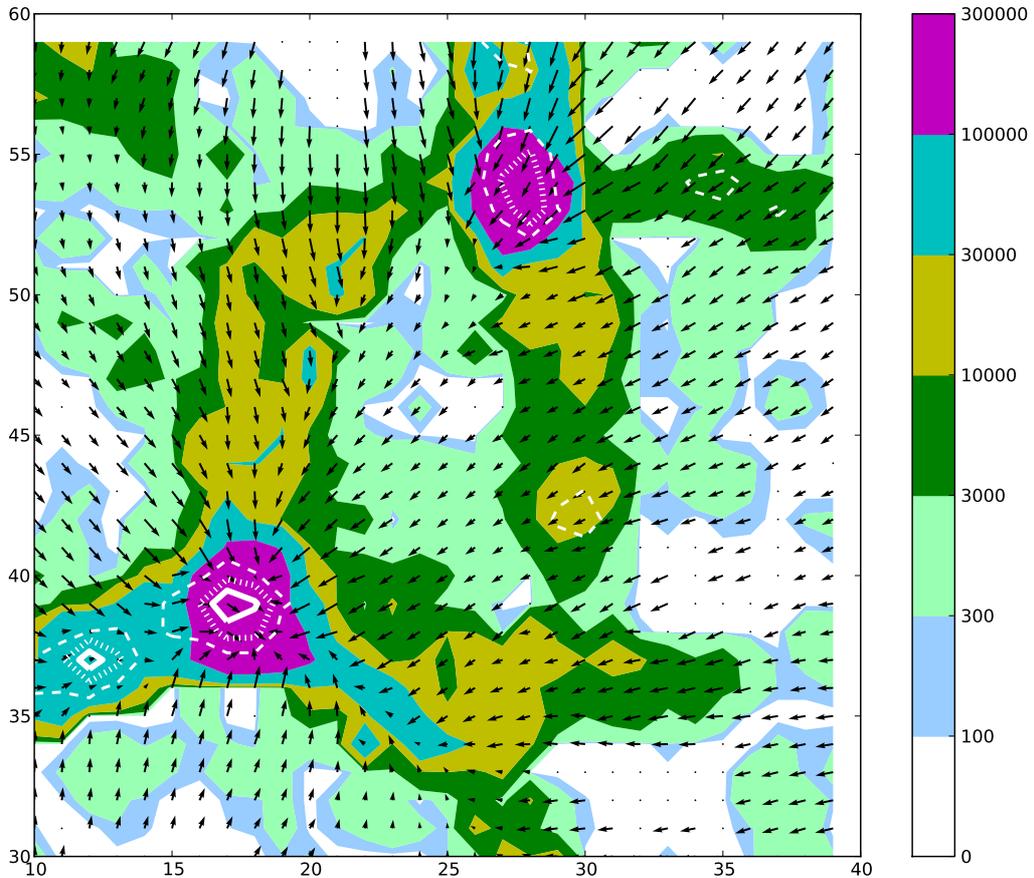}
\caption{The mean velocity  field $\bar{v}$ projected on the slice plane is superimposed on the $\sigma_v^2$ field shown in Fig. \ref{fig:clumps30}.  White contours show peaks
with density above 10, 30 and 60 in the units of the mean density with dashed, dotted and heavy solid lines
respectively. 
}
\label{fig:meanVel-clumps30}
\end{figure}
It is worth stressing that in order to accommodate
the arrow lengths in the figure, their lengths are made proportional to  the square root of 
the magnitude instead of the magnitude. Thus the difference between the fluxes  from three adjacent  
filaments   compared to that from three surrounding voids is considerably more dramatic than
it appears in the figure. 
\begin{figure}
  \centering\includegraphics[width=17.5cm]{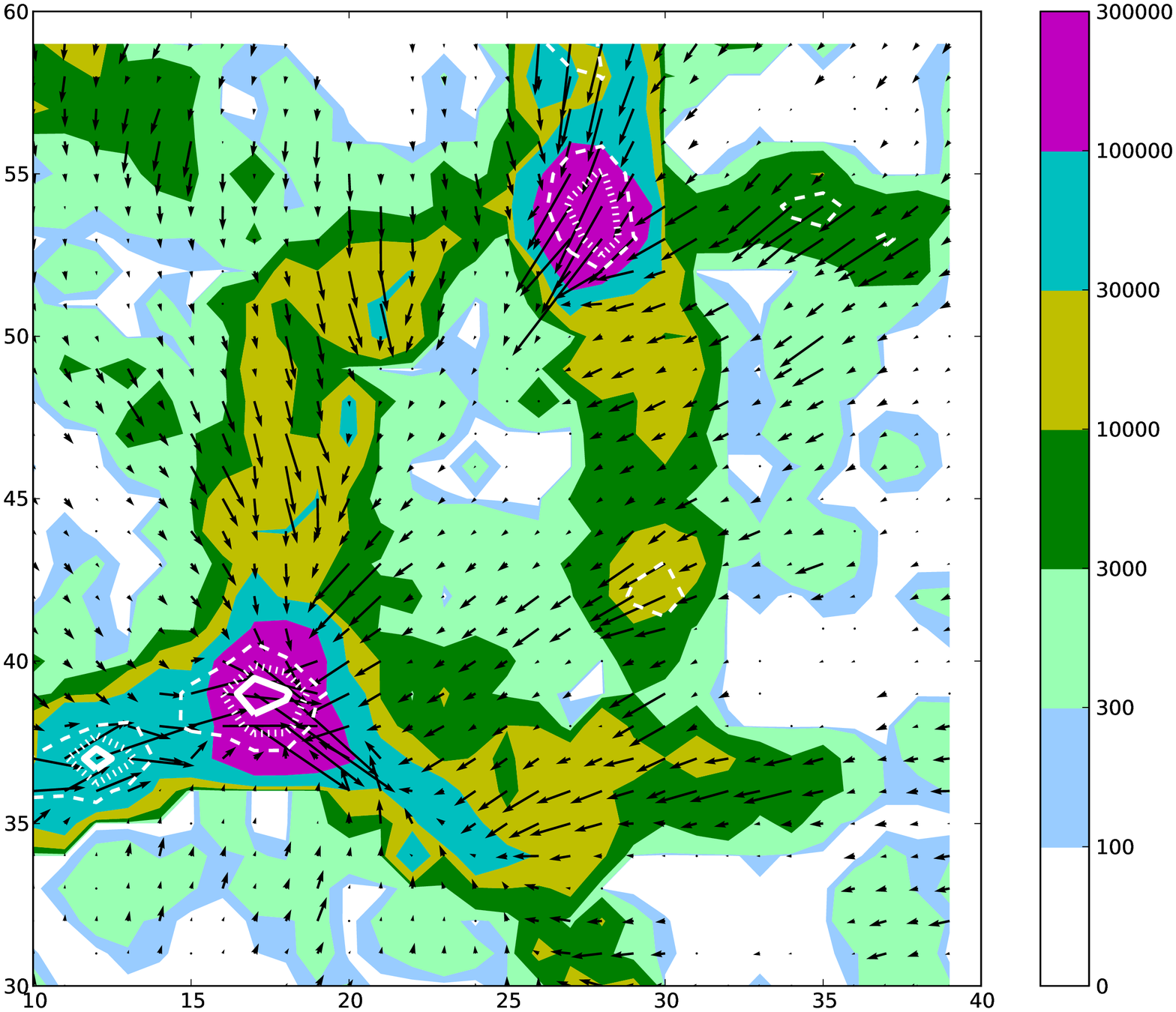}
\caption{The flux of mass $\rho \bar{\bf v}$ is shown by arrows. 
The color scheme and contours are the same as in Fig. \ref{fig:meanVel-clumps30}.
In order to accommodate the length of arrows  their lengths are made proportional
to the square root of the magnitude.
}
\label{fig:mass-flow-clumps30}
\end{figure}

\section{Summary}
\label{sec:summary}
We present a new numerical technique that allows to identify the multi-stream flows, \ie the regions 
where the velocity field is multi-valued. According to \cite{z-70} these regions have 
reached a clear physical threshold to be identified as being in strong nonlinear regime but 
most of them are not virialized.  The multi-stream flows form the web overall very
similar by appearance to the web identified in the density field.  At the same time many
features are very different both in low and high density environments as seen in Fig. \ref{fig:void30} and
\ref{fig:clumps30}. These include the difference of the shapes of the contours around peaks,
the lack of monotonic  relation between the heights of peaks in the $\rho$ and $\sigma_v^2$
fields. Although one can see a positive correlation between the
density and  the  \sv  that determines the multi-stream flows in Fig. \ref{fig:hist2d},
the spread of points along both axes ($\rho$ and \sv) is very large suggesting that these parameters
are quite independent. Therefore \sv may serve 
as a second parameter characterizing the dynamical environment in the nonlinear density peaks
before they reached virial equilibrium as well as for filaments and pancakes.
The method for computing \sv is localized in  Eulerian coordinates and is very easy to implement. 

Ideally \sv is identically equal to zero everywhere where the velocity is singled-valued but the 
simplest numerical implementation  based on the ordinary CIC method generates noise.
Although it is seemingly not posing a serious problem other schemes like  TSC (triangular shape
density cloud)  may be less noisy. This will be checked an reported in the following publications.

The field \sv naturally defines the physical condition for voids as the regions devoid of
multi-stream flows. The identification of the multi-stream flows of course depends 
on the resolution (in particular the mass resolution) of the simulation, 
but this is the fundamental limitation for all fields obtained in the simulations. 
The \sv field by no means is worse than commonly used density field in this respect.
This condition is physically more clear than one based  on an essentially arbitrary density threshold.
It is worth stressing that about 22\% of the total volume has nonzero density but no multi-stream flows.
For the resolution of the \L CDM simulation used here ($\approx$1 \hm1 Mpc) 
we find  the volume fraction in  voids is at least about  one third of the total volume. 
This is considerably greater than at the percolation
transition evaluated for a similar simulation in \cite{sh-etal-04}, therefore the volume devoid
of multi-stream flows consists predominantly of one connected region.

The mean velocity field $\bar{v_i}$ defined by eq. \ref{eq:mean-vel} shows the mass flow along
the filaments toward the clumps in agreement with the prediction of the adhesion approximation 
\citep{gur-etal-89}. Some filaments seem to move predominantly in the transverse direction as a whole 
(see Fig. \ref{fig:meanVel-clumps30}). The mean velocity field seems to be quite smooth without
significance convergence in these cases. A somewhat similar observation can be made 
concerning the clumps. In some of them as in one at (17, 39) in 
Fig. \ref{fig:meanVel-clumps30} and \ref{fig:mass-flow-clumps30} the velocity field seems to
converge significantly stronger than in the other similar clump at (28, 54) in the same figures
where the velocity field looks significantly more steady. 

The filaments with lengths over 50 \hm1 are shown to be the multi-stream flows 
in a qualitative agreement with both the truncated
Zeldovich approximation \citep{col-etal-93} and the adhesion approximation \citep{gur-etal-89}.

Summarizing, we would like to stress the importance of the study of dynamics in filaments and
pancakes, which are highly nonlinear but unvirialized dynamical systems. The suggested technique
may be a useful tool to serve this purpose.
\\

{\bf Acknowledgments}
The author is greateful to Salman Habib  and  Katrin Heitmann for providing the results of N-body
simulation.


\begin{thebibliography}{99}

\bibitem[\protect\citeauthoryear{{Arag\'{o}n-Calvo} et al.} {2007}] {ara-etal-07} 
{Arag\'{o}n-Calvo} M A, Jones B J T, van de Weygaert  R, van der Hulst  J M, 
{\it The multiscale morphology filter: identifying and extracting spatial
patterns in the galaxy distribution},
2007 Astron. Astrophys.  {\bf 474}  315

\bibitem[\protect\citeauthoryear{{Arag\'{o}n-Calvo} et al.} {2010}] {ara-etal-10} 
Arag\'{o}n-Calvo M A, Platen E, van de Weygaert R, Szalay A S, 
{\it The spine of the cosmic web},
2010 Astrophys. J. {\bf 723} 364

\bibitem[\protect\citeauthoryear{{Arnold}, {Shandarin} \& {Zeldovich}} {1982}] {a-sh-z-82} 
{Arnold}  V I, {Shandarin}  S F,  {Zel'dovich}  Ya B, 
{\it The large-scale structure of the universe I. general properties. One- and two-dimensional models}
1982  Geophys. Astrophys. Fluid Dynamics 
  {\bf 20}  111
  
\bibitem[\protect\citeauthoryear{{Bardeen} et al.} {1986}]{bar-etal-86}  
Bardeen  J, Bond  J R, Kaiser  N, Szalay  A S,
{\it The statistics of peaks of Gaussian random fields},
 1986  Astrophys. J.  {\bf 304}  15
   
\bibitem[\protect\citeauthoryear{{Bond}, {Kofman} \& {Pogosyan}} {1996}]{bon-etal-96} 
Bond  J R, Kofman  L,  Pogosyan  D,
{\it How filaments of galaxies are woven into the cosmic web},
  1996  Nature {\bf 380}  603
  
\bibitem[\protect\citeauthoryear{{Bond} et al.} {2010b}]{bon-etal-10b}  
Bond N A, Strauss M A, Cen R,
{\it Crawling the cosmic network: exploring the morphology of structure in the galaxy distribution},
2010 Mon. Not. Roy. Astron. Soc. {\bf 406} 1365

\bibitem[\protect\citeauthoryear{{Bond} et al.} {2010a}]{bon-etal-10a}  
Bond N A, Strauss M A, Cen R,
{\it Crawling the cosmic network: identifying and quantifying filamentary structure},
2010 Mon. Not. Roy. Astron. Soc. [arXiv:1003.3237]


\bibitem[\protect\citeauthoryear{{Buchert}} {1989}]{buc-89}
Buchert T, 
{\it Lighting up pancakes - Towards a theory of galaxy formation},
1989   Rev. Mod. Astr. {\bf 2}  267

\bibitem[\protect\citeauthoryear{Coles, Melott \& Shandarin}{1993}]{col-etal-93}
Coles  P, Melott  A, Shandarin  S F, 
{\it Testing approximations for non-linear gravitational clustering},
1993  Mon. Not. Roy. Astron. Soc.  {\bf 260}  765

\bibitem[\protect\citeauthoryear{{Doroshkevich} \& {Shandarin}} {1978a}]{dor-sh-78a} 
Doroshkevich  A G, Shandarin  S F,
{\it A statistical approach to the theory of galaxy formation},
  1978  Sov. Astron.  {\bf 22}  653

\bibitem[\protect\citeauthoryear{{Doroshkevich} \& {Shandarin}} {1978b}]{dor-sh-78b} 
Doroshkevich  A G, Shandarin  S F,
{\it A mean density and a correlation function of rich clusters: theory and observations},
  1978 Mon. Not. Roy. Astron. Soc. {\bf 182} 27

\bibitem[\protect\citeauthoryear{{Doroshkevich} et al.} {1980}]{dor-etal-80}
Doroshkevich A G, Kotok E V, Novikov I D, Polyudov A N, Shandarin S F, Sigov Yu S,
{\it Two-dimensional simulation of the gravitational system dynamics and formation of the
large-scale structure of the universe}, 
1980 Mon. Not. Roy. Astron. Soc.  {\bf 192} 321

\bibitem[\protect\citeauthoryear{{Fillmore} \& {Goldreich}} {1984}]{fil-gol-84}
Fillmore  J A,  Goldreich  P,
{\it Self-similar gravitaional collapse in an expanding universe},
1984  Astrophys. J.  {\bf 281}  1 

\bibitem[\protect\citeauthoryear{{Forero-Romero} et al.} {2009}]{for-rom-etal-09}
Forero-Romero  J E, Hoffman  Y, Gottl\"{o}ber  S, Klypin  A, Yepes  G,
{\it A dynamical classification of the cosmic web},
2009  Mon. Not. Roy. Astron. Soc. {\bf 396} 1815 

\bibitem[\protect\citeauthoryear{{Gurbatov}, {Saichev} \& {Shandarin}}{1989}]{gur-etal-89} 
{Gurbatov}  S  N, {Saichev}  A I,  {Shandarin}  S  F,
{\it The large-scale structure of the universe in the frame of the model equation of non-linear diffusion},
 1989  Mon. Not. Roy. Astron. Soc. {\bf 236}  385

\bibitem[\protect\citeauthoryear{{Habib} et al.}{2009}]{hab-etal-09} 
Habib  S, et al.
{\it Hybrid petacomputing Meets Cosmology: The Roadrunner Universe Project},
 2009  Journal of Physics: Conference Series  {\bf 180}  012019

\bibitem[\protect\citeauthoryear{{Hahn} et al.}{2007}]{hah-etal-07} 
Hahn  O, Porciani  C, Carollo  C M, Dekel  A,
{\it Properties of dark matter halos in clusters, filaments, sheets and voids},
 2007  Mon. Not. Roy. Astron. Soc.  {\bf 375}  489

\bibitem[\protect\citeauthoryear{{Klypin} \& {Shandarin}} {1983}]{kly-sh-83} 
Klypin  A A, Shandarin  S F,
{\it Three-dimensional numerical model of the formation of large-scale structure in the universe},
  1983  Mon. Not. Roy. Astron. Soc.  {\bf 204}  891

\bibitem[\protect\citeauthoryear{{Kofman} et al.} {1992}]{kof-etal-92} 
Kofman  L,  Pogosyan  D, Shandarin  S F, Melott  A L,
{\it Coherent structures in the universe and the adhesion model},
  1992  Astrophys. J. {\bf 393}  437

\bibitem[\protect\citeauthoryear{{Little}, {Weinberg} \& {Park}} {1991}]{lit-etal-91}  
Little  B, Weinberg  D, Park  C,  
{\it Primordial fluctuations and non-linear structure},
1991  Mon. Not. Roy. Astron. Soc.  {\bf 253}  295

\bibitem[\protect\citeauthoryear{{Maciejewski } et al.} {2010}]{mac-etal-10} 
Maciejewski M, Vogelsberger M, White S D M, Springel V,
{\it Bound and unbound substructures in Galaxy-scale Dark
Matter haloes}, 2010,  arXiv:1010.2491

\bibitem[\protect\citeauthoryear{{Melott} \& {Shandarin}} {1989}]{mel-sh-89}  
Melott  A L, Shandarin  S F,
{\it Gravitational instability with high resolution},
1989  Astrophys. J.  {\bf 343}  26
  
\bibitem[\protect\citeauthoryear{{Melott}, {Shandarin} \& {Weinberg}} {1994}]{mel-etal-94}  
Melott  A L, Shandarin  S F, Weinberg  D,
{\it A test of the adhesion approximation for gravitational clustering},
1994  Astrophys. J.   {\bf 428}  28
  
\bibitem[\protect\citeauthoryear{{Novikov} et al.} {2006}]{nov-etal-06}
Novikov D, Colombi S, Dor\'{e} O., 
{\it Skeleton as a probe of the cosmic web: the two-dimensional case},
2006  Mon. Not. Roy. Astron. Soc.  {\bf 366} 1201

\bibitem[\protect\citeauthoryear{{Peebles}} {1980}]{pee-80} 
Peebles  P J E , {\it Large Scale Structure of the Universe},  1970  , Prinston Univ. Press

\bibitem[\protect\citeauthoryear{{Platen} et al.} {2007}]{pla-etal-07}
Platen E, van de Weygaert R, Jones B J T., 
{\it A cosmic watershed: the WVF void detection technique},
2007 Mon. Not. Roy. Astron. Soc.  {\bf 380} 551

\bibitem[\protect\citeauthoryear{{Pope}, et al.} {2010}]{pop-etal-10} 
Pope  A, Habib  S, Lukic  Z,  Daniel  D, Fasel  P, Desai  N, Heitmann  K,
{\it The accelerated Universe},
  2010  Computing in Science and Engineering {\bf 12}  17
  

\bibitem[\protect\citeauthoryear{{Shandarin}, {Sheth} \& {Sahni}}{2004}] {sh-etal-04} 
{Shandarin}  S F, {Sheth}  J, Sahni  V,
{\it Morphology of the supercluster-void network in \L CDM cosmology},
  2004 Mon. Not. Roy. Astron. Soc. {\bf 353} 162

\bibitem[\protect\citeauthoryear{{Shandarin} \& {Zeldovich}}{1989}] {sh-z-89} 
{Shandarin}  S F, {Zeldovich}  Ya B, 
{\it The large-scale structure of the universe: Turbulence, intermittency, structures 
in a self-gravitating medium},
1989  Rev.  Mod. Phys.  {\bf 61} 185

\bibitem[\protect\citeauthoryear{{Weinberg} \& {Gunn}}{1990a}] {wei-gun-90a} 
Weinberg  D H, Gunn  J,
{\it Simulation of deep redshift surveys},
  1990 Astrophys. J. {\bf 325} L25

\bibitem[\protect\citeauthoryear{{Weinberg} \& {Gunn}}{1990b}] {wei-gun-90b}
Weinberg  D H, Gunn  J,
{\it Large-scale structure and the adhesion approximation},
  1990   Mon. Not. Roy. Astron. Soc.  {\bf 247} 260

\bibitem[\protect\citeauthoryear{{Zeldovich}}{1970}]{z-70} 
{Zeldovich}  Ya B, 
{\it Gravitaional instability: An approximate theory for large density perturbations},
1970  Astron. Astrophys.  {\bf 5} 84 
  
\end{thebibliography}
\end{document}